# QUANTUM TUNNELING IN THE TIME PERIODIC DOUBLE WELL POTENTIAL

V.I. DUBINKO, O.S. MAZMANISHVILI

National Science Center "Kharkiv Institute of Physics & Technology" NASU, 61108, Kharkiv, Ukraine

We present the mathematical model and numerical calculation results for the tunneling of the wave function in a time-periodic double-well potential. The biquadratic potential of a double-well form is used. Based on a mathematical model of the time evolution of the wave function, a numerical algorithm has been developed and a program has been created for solving the Schrödinger equation, which describes the time evolution of the particle's wave function. As a result of numerical modeling, modulation regimes have been obtained at which tunneling took place. In a stationary double well potential, no tunneling was observed. On the contrary, in the time periodic potential, tunneling was observed, the rate of which depended on the modulation frequency. Adjusting the modulation parameters, it is possible to control the particle wave function tunneling rate.

**Introduction**

Quantum tunneling is a quantum mechanical phenomenon in which an object such as an electron or atom passes through a potential energy barrier that, according to classical mechanics, should not be passable due to the object not having sufficient energy to pass or surmount the barrier. Tunneling plays an essential role in physical phenomena such as nuclear fusion and alfa radioactive decay of atomic nuclei. Solving the Schrodinger equation, the time evolution of a known wave function can be deduced. The square of the absolute value of this wave function is directly related to the probability distribution of the particle positions, which describes the probability that the particles would be measured at those positions.

Double well potential is commonly used for calculation of the probability distribution of the particle between two neighboring wells separated by the potential barrier, such as the Coulon repulsion. The tunneling rate depends exponentially on the barrier height and width. So that it becomes negligibly low if the particle energy is much lower than the potential height. This is true for the stationary potential barrier. But one may be curious as to what happens if the potential barrier is not fixed for all time but itself varies with time [1]. In the works [2, 3], it was shown that the extremely small probability of tunneling through an almost classical potential barrier may become not small under the action of the specially adapted non-stationary signal which selects the certain particle energy ER. For particle energies close to this value, the tunneling rate is not small during a finite interval of time and has a very sharp peak at the energy ER. After entering inside the barrier, the particle emits electromagnetic quanta and exits the barrier with a lower energy. The signal amplitude can be much less compared to the field of the static barrier. This phenomenon can be called the Euclidean resonance since the under-barrier motion occurs in imaginary time.

To our knowledge, in all works dealing with nonstationary potential barriers, it was assumed that the *barrier height* was fluctuating with time. In contrast to that, we will consider tunneling in the double well potential, where the *width of the wells* fluctuates periodically in time.

**1. Schrodinger equation**

$$i\hbar \frac{\partial}{\partial t}\psi(x,t) = -\frac{\hbar^2}{2m}\frac{\partial^2}{\partial x^2}\psi(x,t) + V(x,t)\psi(x,t), \quad (1)$$

where $\hbar$ is the Planck constant, $m$ is the particle mass, $\psi(x,t)$ is the waive function and $V(x,t)$ is the double well potential, which depends on time as follows:

$$V(x,t) = \frac{\hbar\omega_0}{2}\left[a(t)\left(\frac{x}{\xi}\right)^4 - b(t)\left(\frac{x}{\xi}\right)^2 + \frac{b^2(t)}{4a(t)}\right], \quad (2)$$

$$a(t) = \frac{\alpha - \beta\cos(\varepsilon\omega_0 t)}{2\sqrt{\alpha}}, \quad b(t) = \frac{\sqrt{\alpha - \beta\cos(\varepsilon\omega_0 t)}}{2\sqrt{\alpha}},$$

where $\omega_0$ is the eigenfrequency of the parabolic potential near the well bottom, where the ratio $\beta/\alpha \ll 1$, $\xi$ is proportional to the root-mean-square fluctuation (standard deviation) of zero-point oscillations of a linear oscillator in the stationary case $\delta x = \sqrt{\langle(\Delta x)^2\rangle} = \sqrt{\langle x^2\rangle - \langle x\rangle^2}$, $\xi = \sqrt{\hbar/m\omega_0} = \sqrt{2}\delta x_0$.

Potential parameters change in time so that the minimum and maximum energy are constant, while time periodic change of the minimum coordinates take place.

The potential height $\Delta V$ is *constant* and the minimum positions are determined as follows:

$$\frac{\Delta V(t)}{(\hbar\omega_0/2)} = \frac{b^2(t)}{4a(t)} = \frac{1}{8\sqrt{\alpha}} = const, \quad (3)$$

$$\frac{x_{min}^{(\pm)}(t)}{\xi} = \pm\sqrt{\frac{b(t)}{2a(t)}} = \pm\frac{1}{\sqrt{2}}\frac{1}{\sqrt[4]{\alpha - \beta\cos(\varepsilon\omega_0 t)}}.$$

In case of $\beta = 0$ (stationary potential) one has $x_{min}^{(\pm)}(t)/\xi = \pm(4\alpha)^{-1/4}$. In dimensionless variables, one has:

$$\tau = \frac{\omega_0 t}{2}, \quad \tilde{x} = \frac{x}{\xi} = \frac{x}{\sqrt{\hbar/m\omega_0}},$$

$$u(\tilde{x},\tau) = \frac{V(x/\xi\omega_0\tau/2)}{(\hbar\omega_0/2)} = \frac{V(\tilde{x},\tau)}{(\hbar\omega_0/2)}, \quad (1)$$

$$i\frac{\partial}{\partial\tau}\psi = -\frac{\partial^2}{\partial\tilde{x}^2}\psi + u(\tilde{x},\tau)\psi, \quad (2)$$

$$u(\tilde{x},\tau) = a(\tau)\tilde{x}^4 - b(\tau)\tilde{x}^2, \quad (6)$$

$$a(\tau) = \frac{\alpha - \beta\cos(\varepsilon\tau)}{2\sqrt{\alpha}}, \quad b(\tau) = \frac{\sqrt{\alpha - \beta\cos(\varepsilon\tau)}}{2\sqrt{\alpha}}.$$

We chose the following parameters $\alpha = 0.0005$, $\beta = 0.0001$, $\omega_0 = 1$. The coordinates of the minimum potential are $x_{\min}^{(\pm)(t)}/\xi(4\alpha)^{-1/4}$. Parameter $\varepsilon$ was chosen to be 0.0, 2.0 and 1.7. `Numeric calculations of the total energy and of the particle energy density were obtained:

$$E(\psi) = \xi\int_{-\infty}^{\infty} w(\tilde{x},\tau)d\tilde{x},$$

$$w(\tilde{x},\tau) = \frac{\hbar\omega_0}{2}\left\{\left|\frac{\partial\psi}{\partial\tilde{x}}\right|^2 + u(\tilde{x},\tau)|\psi|^2\right\}. \quad (3)$$

The coordinates of the particle were measured in $\xi$, the time was measured in $T = 2\pi/\omega_0$, energy was measured in the zero-point energy $(\hbar\omega_0/2)$.

Fig. 1 shows the potential shape at the beginning $\tau = 0$. Probability density is shown as $p(\tilde{x},0) = |\psi(\tilde{x},0)|$

$$\psi(\tilde{x},0) = \frac{1}{\sqrt[4]{\pi\xi^2}}\exp\left(-\frac{(\tilde{x}-\tilde{x}_{\min})^2}{2}\right). \quad (8)$$

To solve the eq. (5) with initial condition (8) we used the MathCAD program based on the Runge-Kutt method and the program for partial differential equations.

### 2. Stationary potential

In a stationary state (Fig. 2) the probability density $p(\tilde{x},\tau) = |\psi(\tilde{x},\tau)|^2$ is in the left well for the time interval $0 \leq \tau \leq 2500$.

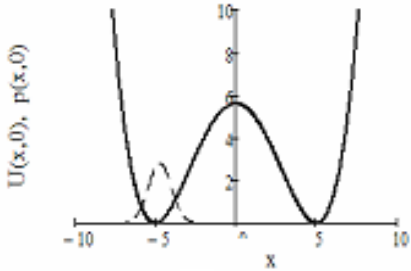

Figure 1. Initial potential (solid curve) and the probability density (dashed curve) of the wave function

Small fluctuations of the probability density are due to the difference of the left well from the parabolic one. The barrier height is $h_b = 1/8\sqrt{\alpha} = 5.590$, while the energy of the particle at the beginning $E = 0.456$.

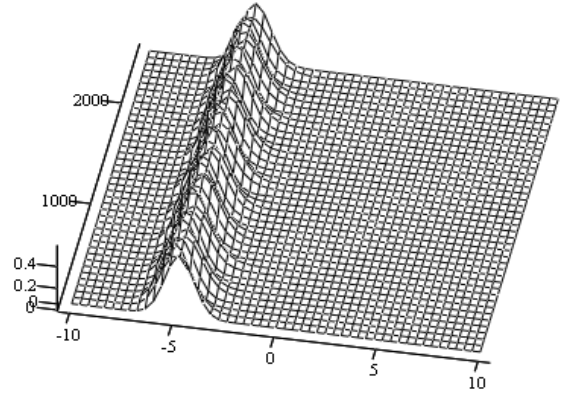

Figure 2. Evolution of the probability density $p(\tilde{x},\tau)$ at $0 \leq \tau \leq 2500$

### 3. Time periodic potential

Fig. 3 shows the quantum tunneling at $\varepsilon = 2.0$, $0 \leq \tau \leq 2500$.

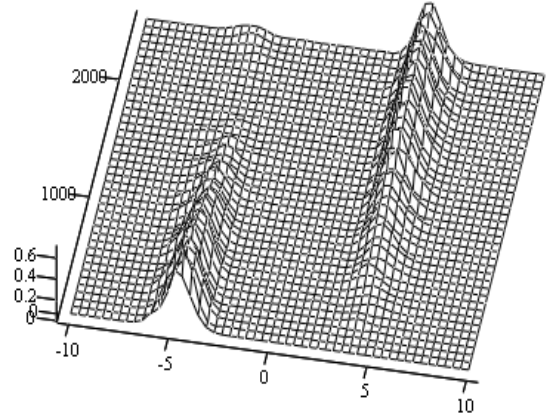

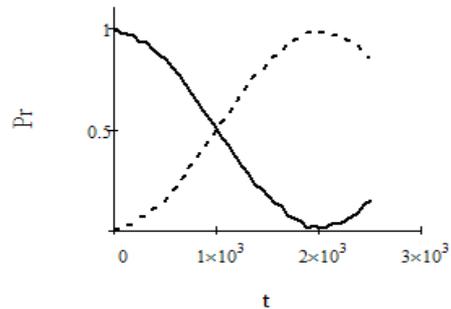

Figure 3. Tunneling at $\varepsilon = 2.0$.
Top: evolution of the probability density, bottom: dynamics of the mean coordinate of the particle

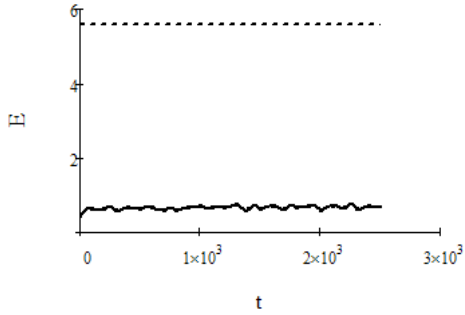

Figure. 4. Tunneling at $\varepsilon = 2.0$.
Top: Time evolution of the probability density – in the left well (solid curve) and in the right well (dashed curve), bottom: energy of the particle (solid curve) and the barrier height $h_b = 5.590$ (dashed curve)

One can see that in the time periodic potential there is tunneling from the left well to the right well and from the right well to the left well. We searched for the frequency parameter $\varepsilon$, at which the tunneling rate increased.

Fig. 5 and 6 shows the tunneling rate at $0 \leq \tau \leq 2500$, $\varepsilon = 1.7$. One can see that in this case, the tunneling rate increases up to seven cycles.

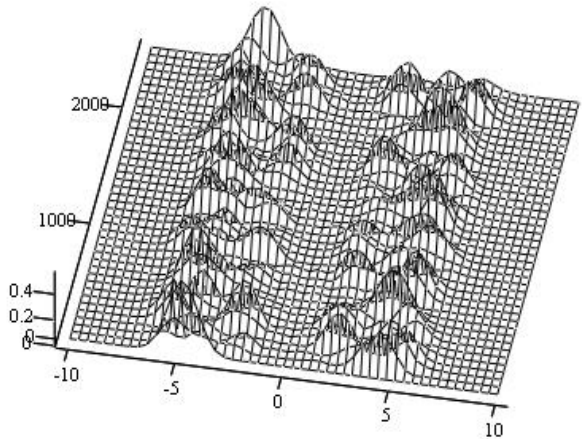

Figure 5a. Tunneling at $\varepsilon = 1.7$.
Evolution of the probability density

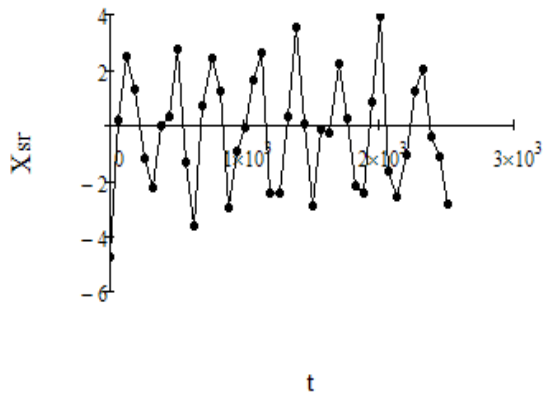

Figure 5b. Tunneling at $\varepsilon = 1.7$.
Dynamics of the mean coordinate of the particle

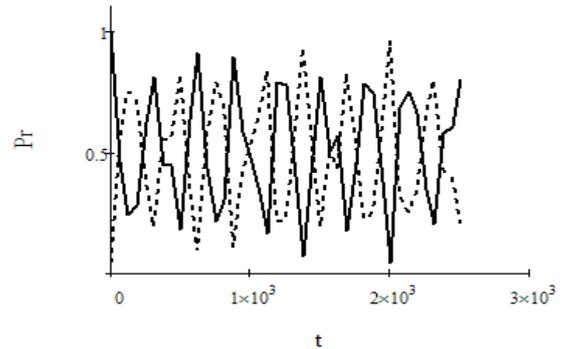

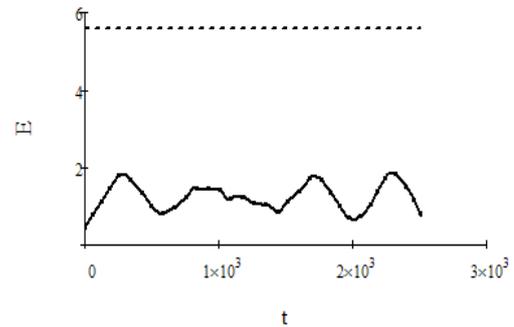

Figure 6. $\varepsilon = 1.7$.
Top: evolution of the probability density in the left well (solid curve) and in the right well (dashed curve), bottom: energy of the particle (solid curve) and the barrier height $h_b = 5.590$ (dashed curve)

### 4. Discussion

In the present paper, we presented a numerical solution of the Schrodinger equation for a particle in a *non-stationary* double well potential, the width of which was driven *time-periodically*. We have shown that the rate of tunneling of the particle through the potential barrier separating the wells can be enhanced by *orders of magnitude* with increasing number of driving periods. This effect is novel, and it differs qualitatively from a well-studied effect of resonance tunneling [2-3], a.k.a. Euclidean resonance (an easy penetration through a classical nonstationary barrier due to an under-barrier interference). In the latter case, the tunneling rate has a sharp peak as a function of the particle energy when it is close to the certain *resonant value* defined by the nonstationary field. Therefore, it requires a very specific parametrization of the tunneling conditions. In contrast to that, the time-periodic driving of the potential wells considered above, results, first, in a sharp and continuous (not quantum) increase of the amplitude and energy of zero-point vibrations [4], which in its turn increases the tunneling rate.

### 5. Summary

The mathematical model and numerical calculation results for the tunneling of the wave function in a time-periodic double-well potential was presented. Based on a mathematical model of the time evolution of the wave function, a numerical algorithm has been developed and a program has been created for solving the Schrödinger equation, which describes the time evolution of the particle's wave function. As a result of numerical modeling, modulation regimes have been obtained at which tunneling took place. In a stationary double well potential no tunneling was observed. On the contrary, in the time periodic potential, tunneling was observed with a rate depending on the modulation frequency. Adjusting the modulation parameters, it is possible to control the particle wave function tunneling rate.

**References**

1. Pranab S. Tunneling through a time-dependent barrier – a numerical study// *Pramana Journal of Physics* - 2000.- V. 54.- No 3.- pp. 385–392.
2. Ivlev B.I. The Euclidean resonance and quantum tunneling// ArXiv: quant-/ph/0202145v1-2002.
3. Palomares-Báez J.P., B. Ivlev B., Rodríguez-López J.L. Enhanced tunneling through nonstationary barriers// *Phys. Rev. A* – 2007.- V. **76**, pp. 052103 (1-8).
4. Dubinko V.I. Chemical and nuclear catalysis driven by localized anharmonic vibrations / V.I. Dubinko, D.V. Laptev// *Letters on materials.* – 2016.-6(1)-pp.-16-21.

## QUANTUM TUNNELING IN THE TIME PERIODIC DOUBLE WELL POTENTIAL

V.I. DUBINKO, O.S. MAZMANISHVILI

National Science Center "Kharkiv Institute of Physics & Technology" NASU, 61108, Kharkiv, Ukraine

We present the mathematical model and numerical calculation results for the tunneling of the wave function in a time-periodic double-well potential. The biquadratic potential of a double-well form is used. Based on a mathematical model of the time evolution of the wave function, a numerical algorithm has been developed and a program has been created for solving the Schrödinger equation, which describes the time evolution of the particle's wave function. As a result of numerical modeling, modulation regimes have been obtained at which tunneling took place. In a stationary double well potential, no tunneling was observed. On the contrary, in the time periodic potential, tunneling was observed, the rate of which depended on the modulation frequency. Adjusting the modulation parameters, it is possible to control the tunneling rate of the particle wave function.

## КВАНТОВЕ ТУНЕЛЮВАННЯ В ПЕРІОДИЧНОМУ В ЧАСІ ПОТЕНЦІАЛІ ПОДВІЙНОЇ ЯМИ

В.І. ДУБІНКО, О.С. МАЗМАНІШВІЛІ

Національний науковий центр "Харківський фізико-технічний інститут" НАНУ, 61108, Харків, Україна

Наведено математичну модель і результати чисельних розрахунків для тунелювання хвильової функції в періодичному в часі двоямному потенціалі. Використовується біквадратичний потенціал двоямної форми. На основі математичної моделі еволюції хвильової функції в часі розроблено чисельний алгоритм і створено програму для розв'язування рівняння Шредінгера, яке описує еволюцію хвильової функції частинки в часі. В результаті чисельного моделювання отримано режими модуляції, при яких відбувалося тунелювання. У стаціонарному потенціалі подвійної ями тунелювання не спостерігалося. Навпаки, у періодичному потенціалі спостерігалося тунелювання, швидкість якого залежала від частоти модуляції. Налаштовуючи параметри модуляції, можна контролювати швидкість тунелювання хвильової функції частинок.